\documentclass [12pt]{cernart}
\usepackage{times}
\usepackage{epsfig}
\usepackage{amssymb}
\usepackage{cite}
\begin{document}
%
\newcommand{\PK}{\ensuremath{\mathrm{K}}}
\newcommand{\PaK}{\ensuremath{\mathrm{\overline{K}}}}
\newcommand{\PKz}{\ensuremath{\mathrm{K^0}}}
\newcommand{\PaKz}{\ensuremath{\mathrm{\overline{K}\mbox{}^0}}}
\newcommand{\PKL}{\ensuremath{\mathrm{K_L}}}
\newcommand{\PKS}{\ensuremath{\mathrm{K_S}}}
\newcommand{\PKLS}{\ensuremath{\mathrm{K_{L,S}}}}
\newcommand{\PKp}{\ensuremath{\mathrm{K^+}}}
\newcommand{\PKm}{\ensuremath{\mathrm{K^-}}}
\newcommand{\PKpm}{\ensuremath{\mathrm{K^\pm}}}
\newcommand{\PKst}{\ensuremath{\mathrm{K^*}}}
\newcommand{\PKstz}{\ensuremath{\mathrm{K^*(892)^0}}}
\newcommand{\Pgp}{\ensuremath{\mathrm{\pi}}}
\newcommand{\Pgpm}{\ensuremath{\mathrm{\pi^-}}}
\newcommand{\Pgpp}{\ensuremath{\mathrm{\pi^+}}}
\newcommand{\Pgm}{\ensuremath{\mathrm{\mu}}}
\newcommand{\Pgn}{\ensuremath{\mathrm{\nu}}}
\newcommand{\Pgf}{\ensuremath{\mathrm{\phi}}}
\newcommand{\Pp}{\ensuremath{\mathrm{p}}}
\newcommand{\PN}{\ensuremath{\mathrm{N}}}
\newcommand{\Pap}{\ensuremath{\mathrm{\overline{p}}}}
\newcommand{\PgL}{\ensuremath{\mathrm{\Lambda}}}
\newcommand{\PC}{\ensuremath{\mathrm{C}}}
\newcommand{\PCtw}{\ensuremath{\mathrm{^{12}C}}}
%
\newcommand{\re}{\mathrm{Re}}
\newcommand{\im}{\mathrm{Im}}
\renewcommand{\d}{\mathrm d}
\newcommand{\e}{\mathrm e}
\newcommand{\fm}{f^{(-)}}
\newcommand{\refm}{\mathrm{Re}\;f^{(-)}}
\newcommand{\imfm}{\mathrm{Im}\;f^{(-)}}
\newcommand{\redf}{\mathrm{Re}\;\Delta f}
\newcommand{\imdf}{\mathrm{Im}\;\Delta f}
\enlargethispage{2\baselineskip}
\begin{titlepage}
\docnum{CERN--EP/99--34}
\enlargethispage{2\baselineskip} 
\date{February 26, 1999} 
\title{Dispersion relation analysis of the neutral kaon regeneration
       amplitude in carbon}
\vspace*{-5mm}
\collaboration{CPLEAR Collaboration \\[0.5\baselineskip]}

\begin{Authlist}
A.~Angelopoulos\Iref{a1}, 
A.~Apostolakis\Iref{a1},
E.~Aslanides\Iref{a11}, 
G.~Backenstoss\Iref{a2}, 
P.~Bargassa\Iref{a13},
O.~Behnke\Iref{a17}, 
A.~Benelli\Iref{a2}, 
V.~Bertin\Iref{a11}, 
F.~Blanc\IIref{a7}{a13},
P.~Bloch\Iref{a4}, 
P.~Carlson\Iref{a15}, 
M.~Carroll\Iref{a9},
E.~Cawley\Iref{a9}, 
M.B.~Chertok\Iref{a3},
M.~Danielsson\Iref{a15}, 
M.~Dejardin\Iref{a14},
J.~Derre\Iref{a14}, 
A.~Ealet\Iref{a11}, 
C.~Eleftheriadis\Iref{a16}, 
L.~Faravel\Iref{a7},
W.~Fetscher\Iref{a17}, 
M.~Fidecaro\Iref{a4},
A.~Filip\v ci\v c\Iref{a10}, 
D.~Francis\Iref{a3}, 
J.~Fry\Iref{a9},
E.~Gabathuler\Iref{a9}, 
R.~Gamet\Iref{a9}, 
H.-J.~Gerber\Iref{a17},
A.~Go\Iref{a4}, 
A.~Haselden\Iref{a9},
P.J.~Hayman\Iref{a9}, 
F.~Henry-Couannier\Iref{a11},
R.W.~Hollander\Iref{a6}, 
K.~Jon-And\Iref{a15}, 
P.-R.~Kettle\Iref{a13},
P.~Kokkas\Iref{a4}, 
R.~Kreuger\Iref{a6},
R.~Le Gac\Iref{a11}, 
F.~Leimgruber\Iref{a2}, 
I.~Mandi\' c\Iref{a10}, 
N.~Manthos\Iref{a8},
G.~Marel\Iref{a14}, 
M.~Miku\v z\Iref{a10}, 
J.~Miller\Iref{a3}, 
F.~Montanet\Iref{a11},
A.~Muller\Iref{a14},
T.~Nakada\Iref{a13}, 
B.~Pagels\Iref{a17},
I.~Papadopoulos\Iref{a16},
P.~Pavlopoulos\Iref{a2},
G.~Polivka\Iref{a2}, 
R.~Rickenbach\Iref{a2}, 
B.L.~Roberts\Iref{a3},
T.~Ruf\Iref{a4}, 
M.~Sch\"afer\Iref{a17}, 
L.A.~Schaller\Iref{a7}, 
T.~Schietinger\Iref{a2},
A.~Schopper\Iref{a4}, 
L.~Tauscher\Iref{a2},
C.~Thibault\Iref{a12}, 
F.~Touchard\Iref{a11}, 
C.~Touramanis\Iref{a9},
C.W.E.~Van Eijk\Iref{a6},
S.~Vlachos\Iref{a2}, 
P.~Weber\Iref{a17}, 
O.~Wigger\Iref{a13}, 
M.~Wolter\Iref{a17},
D.~Zavrtanik\Iref{a10},
D.~Zimmerman\Iref{a3}\\[0.5\baselineskip]
and \\[0.5\baselineskip]
M.P.~Locher\Iref{a18},
V.E.~Markushin\Iref{a18}

\end{Authlist}

\vspace{5mm}
\begin{abstract}
We apply a forward dispersion relation to the regeneration amplitude 
for kaon scattering on $\PCtw$ using all available data.
The CPLEAR data at low energies allow the determination of the net contribution  
from the subthreshold region which turns out to be much smaller than earlier 
evaluations, solving a long standing puzzle. 
\end{abstract}
\submitted{(submitted to The European Physical Journal C)}
\Instfoot{a1}{University of Athens, Greece}
\Instfoot{a2}{University of Basle, Switzerland}
\Instfoot{a3}{Boston University, USA}
\Instfoot{a4}{CERN, Geneva, Switzerland}
\Instfoot{a5}{LIP and University of Coimbra, Portugal}
\Instfoot{a6}{Delft University of Technology, Netherlands}
\Instfoot{a7}{Uni\-ver\-sity of Fribourg, Switzerland}
\Instfoot{a8}{Uni\-ver\-sity of Ioannina, Greece}
\Instfoot{a9}{Uni\-ver\-sity of Liverpool, UK}
\Instfoot{a10}{J.~Stefan Inst.\ and Phys.\ Dep., University of Ljubljana, Slovenia}
\Instfoot{a11}{CPPM, IN2P3-CNRS et Universit\'e d'Aix-Marseille II, France}
\Instfoot{a12}{CSNSM, IN2P3-CNRS, Orsay, France}
\Instfoot{a13}{Paul Scherrer Institut (PSI), Villigen, Switzerland}
\Instfoot{a14}{CEA, DSM/DAPNIA, CE-Saclay, France}
\Instfoot{a15}{Royal Institute of Technology, Stockholm, Sweden}
\Instfoot{a16}{University of Thessaloniki, Greece}
\Instfoot{a17}{ETH-IPP Z\"urich, Switzerland}
\Instfoot{a18}{Paul Scherrer Institut (PSI), Theory Group, Villigen, Switzerland} 
%
\end{titlepage}

\section{Introduction} \label{Introduction}

Regeneration of neutral kaons is an excellent tool to study
kaon-nuclear scattering \cite{Gsponer} and of crucial importance 
to experiments measuring CP violation in the kaon system 
\cite{Orsay}. 
  In the latter context, CPLEAR \cite{CPLEAR} has determined 
the kaon regeneration amplitude on $\PC$
in the energy range between 0.56 and 0.9~GeV. 
In the present paper we explore the consequence of these data 
for strong-interaction physics analyzing the kaon-carbon scattering amplitude 
in the classical framework of forward dispersion relations (FDR) using 
in addition all the available information on $\PK\PC$ scattering
\footnote{%
All the data refer to natural carbon, with a $98.9\%$ of $\PCtw$ content, to
which the analysis applies. }:  
the kaon regeneration on carbon 
\cite{Christenson,Boehm,Bott67,Bott69,Carithers75a,Carithers75b,%
Carithers77,Hladky,Albrecht,Roehrig,Schwingenheuer}, 
total cross sections \cite{Bugg,Abrams,Gobbi,Afonasyev}, 
near forward differential cross sections \cite{Gobbi}, 
and the scattering length from kaonic atoms \cite{Backenstoss,Seki}. 
For high energies, we use the Regge model which is well supported experimentally 
\cite{Roehrig, Schwingenheuer}. For the low energy application presented 
here this last aspect is not important. The energy range of the CPLEAR experiment 
is however well suited to obtain information on the strong-interaction 
physics occurring below the elastic threshold, often termed the unphysical 
region. 
   Our results indicate that previous evaluations of FDRs have to be revised 
entirely in this region. Due to the new data situation, the net contribution 
from the unphysical region turns out to be much smaller and a simple coherent 
sum of elementary $\PK\PN$ amplitudes is inadequate and even has the wrong 
sign as will be discussed below.    

\section{Selection and treatment of data} \label{Selection}

We define the regeneration amplitude
as the difference between the nuclear amplitudes of forward 
scattering of \PaKz\ and \PKz\ on \PC: 
\begin{equation}
  \fm(\omega) = f(\PaKz\PC;\omega,\theta=0) - f(\PKz\PC;\omega,\theta=0)
\label{Eqfm}
\end{equation} 
where $\omega$ is the total laboratory energy of the kaon. 
Note that this definition, which we choose for compatibility with earlier  
evaluations of FDR (an alternative definition with an extra factor $1/2$ in 
the r.h.s. of Eq.~\ref{Eqfm} is also used in the literature), 
has the opposite sign when compared to the $\Delta f$ 
of our earlier regeneration analysis \cite{CPLEAR}.

Table~1 summarizes all experimental efforts to determine 
the regeneration amplitude in carbon.
Most regeneration experiments determined modulus and phase of
$\fm$ from \Pgpp\Pgpm\ decay rates behind a regenerator
\cite{CPLEAR, Christenson, Boehm, Bott69, Carithers75b, Carithers77,
 Albrecht, Hladky,  Roehrig, Schwingenheuer}.
Two groups \cite{Bott67, Carithers75a} also obtained a phase 
measurement from the time-dependent charge-asymmetry in semileptonic 
decays, thus independent of the CP violation phase $\phi_{+-}$.

Owing to the isospin symmetry of the \PCtw\ nucleus, \PKpm\ data 
may be used to extract information on the regeneration amplitude,
if charge invariance is assumed:
\begin{equation}
  \fm(\omega) = f(\PKm\PC;\omega,\theta=0) - f(\PKp\PC;\omega,\theta=0)
\end{equation} 
Then, with the use of the optical theorem, the difference between
the total cross sections of \PKm\ and \PKp\
is a measure of the imaginary part of $\fm$
\cite{Bugg, Abrams, Gobbi, Afonasyev} (Table~2).
One of these experiments \cite{Gobbi} also derived values for the real 
part from a measurement of the differential cross section 
$\d\sigma/\d q^2$.

Similarly, the \PKm\PC\ scattering length, $a_0(\PKm\PC)$, 
extracted \cite{Seki} from X-ray transition data obtained with
kaonic carbon \cite{Backenstoss}
may be regarded as a measurement of 
the \PaKz\ scattering amplitude near zero kinetic energy.
The corresponding amplitude for \PKz, having no inelastic channels open  
below threshold, must be purely real such that 
$\imfm(m_\PKz) = -\im\; a_0(\PKm\PC)$.

Where appropriate, the results have been adapted to the current 
world averages for the neutral kaon parameters, such as $\Delta m$, 
$\tau_{\mathrm{S}}$ and $\eta_{+-}$, taken from Ref.~\cite{PDG}.
Here it is important to note that some of these averages, in 
particular $\phi_{+-}$, are themselves influenced by regeneration 
measurements so that care must be taken when employing them to 
correct these very measurements.
Such feedback effects, however, were found to be very small and 
may be neglected in our analysis.

\begin{figure}[b]
\begin{center}
    \mbox{\epsfig{figure=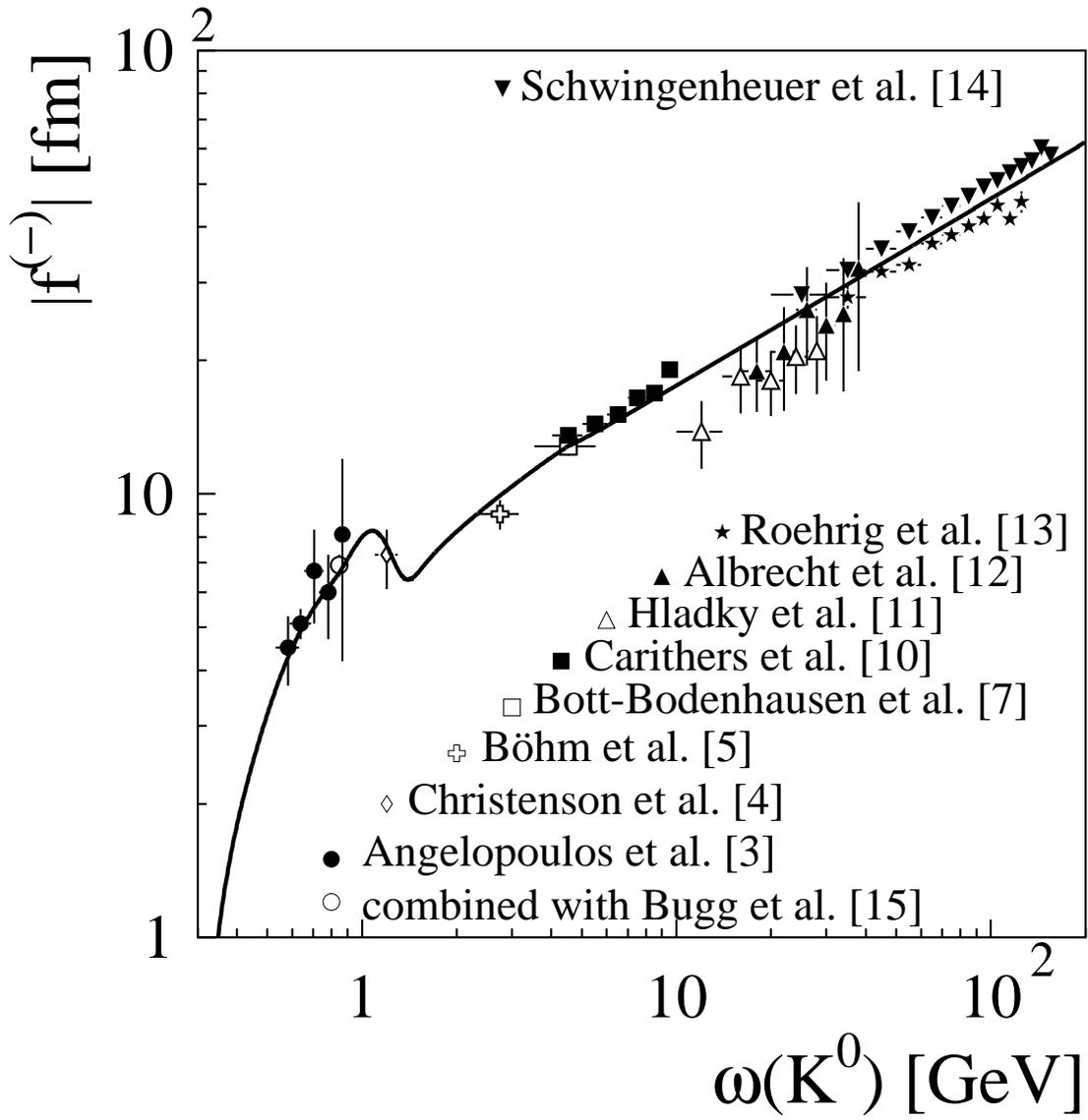,width=16cm}}
\end{center} 
\caption{\label{FigModule}
             Modulus of the regeneration amplitude $\fm$:
             experimental results in comparison with our fit (solid line).
             The data points of Schwingenheuer et al.\ are taken
             from Ref.~\cite{Briere} and are shown just for 
             visualization (only their published value for $\alpha$ 
             enters the fit).
             Note the double-logarithmic scale.}
\end{figure}

\begin{figure}[b]
\begin{center}
   \mbox{\epsfig{figure=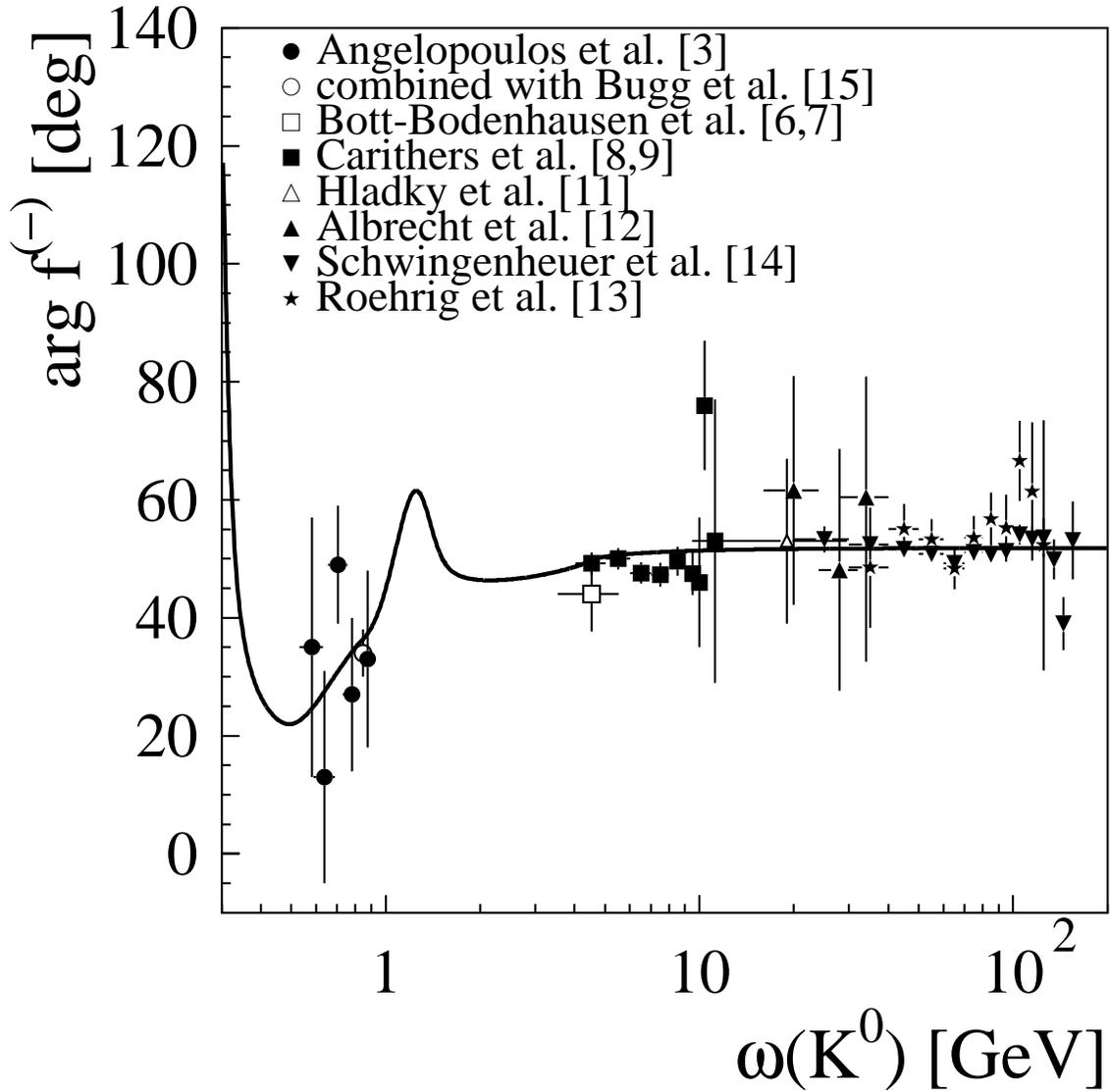,width=16cm}}
\end{center} 
\caption{\label{FigPhase}    
             Phase of the regeneration amplitude $\fm$:
             experimental results in comparison with our fit (solid line).
             The data points of Schwingenheuer et al.\ are taken
             from Ref.~\cite{Briere} and are shown just for 
             visualization (only their published value for $\alpha$ 
             enters the fit).
             The data of Refs.~\cite{Bott67} and \cite{Bott69} as well 
             as those of Refs.~\cite{Carithers75a} and 
             \cite{Carithers75b} have been averaged for clarity.
             Note the logarithmic scale.}
\end{figure}

\begin{figure}[b]
\begin{center}
   \mbox{\epsfig{figure=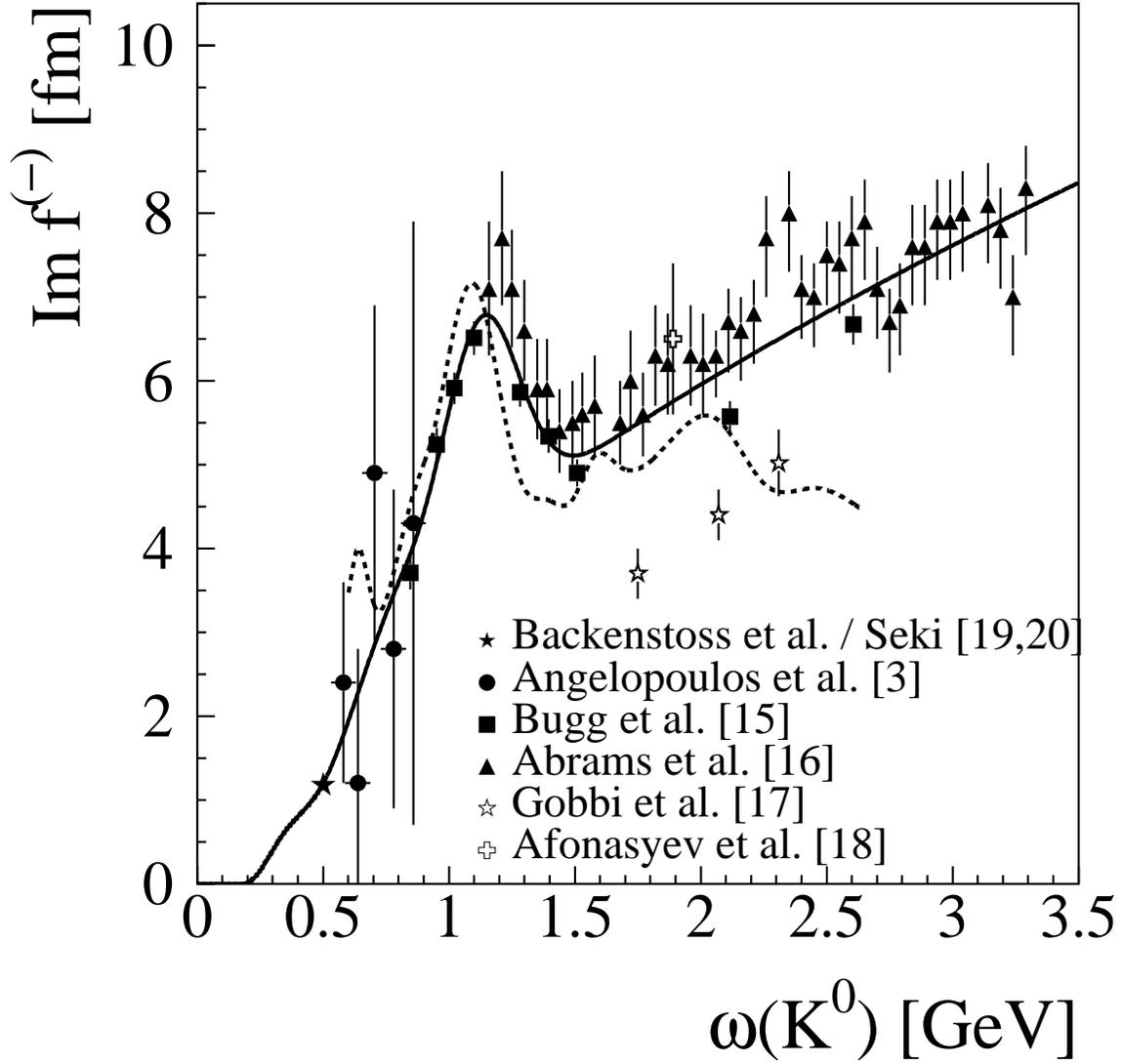,width=16cm}}
\end{center} 
\caption{\label{FigIm} 
             Imaginary part of the regeneration amplitude $\fm$:
             experimental results in comparison with our fit (solid line)
             and the optical model calculation of Ref.~\cite{Eberhard}
             (dashed line).
             The dotted line represents the sub-threshold parameterization
             employed in Ref.~\cite{Arai}.
             The data of Ref.~\cite{Gobbi} are discarded in the fit.
             Note the linear scale.}
\end{figure}

The corrected data are illustrated in Figs.~\ref{FigModule}--
\ref{FigRe}, separately for
measurements of modulus, phase, imaginary and real part of $\fm$.
Figure~\ref{FigIm} reveals a conspicuous systematic disagreement
between the total 
cross section results of Ref.~\cite{Gobbi} and the data of the 
other three experiments at the same energies.
We exclude the data published in Ref.~\cite{Gobbi} from our fit,
imaginary parts as well as real parts, as the latter directly depend
on the former.  For their impact see Section~\ref{FitRes}.

\section{The dispersion relation and its parameterization} \label{DRP}

   In the framework of local relativistic field theories,
analyticity of the amplitude has been proven to follow from local causality
and it is therefore very natural to assume it as well for kaon scattering 
off nuclei.
  We use an unsubtracted dispersion relation as it is readily
derived for antisymmetric amplitudes \cite{Ericson}.
Since the precision of the data does not allow fitting several pole
terms, they are represented by a single effective pole:
\begin{equation} \label{DR}
  \refm(\omega) = \frac{2\omega r}{\omega^2 - \omega_\mathrm{p}^2} +
  \frac{2\omega}{\pi}\mathcal{P}\int_{\omega_\PgL}^{\infty} \d\omega'
  \frac{\imfm(\omega')}{\omega'^2 - \omega^2}
\end{equation}
where $\omega$ denotes the total laboratory energy of the kaon and
$\mathcal{P}$ stands for principal value.
   The effective pole term contains two parameters, position $\omega_\mathrm{p}$ 
and residue $r$. The lowest inelastic threshold $\omega_\PgL = 183.6$ MeV
corresponds to $\Lambda^{11}\!\mathrm{X}$ system at zero energy,
see Fig.~\ref{FigDiag}a.
The integration over the imaginary part starts at $\omega_\PgL$
and extends all the way to infinity (the physical region of \PKz\ and \PaKz\
scattering on \PC\ corresponds to $\omega\geq m_\PKz$).
To carry out the integration, a parameterization of $\imfm$ is
needed over this range.
Guided by the data and theoretical expectations, we have chosen the 
following parameterization:
\begin{itemize}
  \item
    The Regge model with one pole trajectory exchange \cite{Cabibbo} 
    is applied at energies above a certain energy $\omega_\mathrm{k}$, 
    giving 
    \begin{equation}
      \imfm(\omega) = \beta(\omega/\omega_0)^\alpha
      \sin(\frac{\pi}{2}[\alpha+1]).
    \label{EqRegge}
    \end{equation}
    We choose the dimensional parameter  $\omega_0 =$ 1 MeV.
    With the slope parameter $\alpha \approx 0.42$ (see below),
    the rate of convergence of the dispersion integral (\ref{DR})
    at $\omega\to\infty$ is sufficient for reliable calculations
    in the near-threshold region.
  \item 
    In the intermediate energy range around 1 GeV, the total cross 
    section data clearly indicate a resonance. 
    We parameterize it for simplicity as a Gaussian on a background, the shape of 
    which we describe by the same simple power law as employed at
    high energies, but with a different exponent: 
    \begin{equation}
      \imfm(\omega) = b(\omega/\omega_0)^a\sin(\frac{\pi}{2}[a+1])
      + C \exp\left(-\frac{1}{2}
       \left[\frac{\omega-\omega_\mathrm{res}}{\sigma}
                                            \right]^2\right)
    \label{ImfBGpeak}
    \end{equation}
    The transition energy $\omega_\mathrm{k}$ follows from $\alpha$,
    $\beta$, $a$ and $b$ if the function $\imfm(\omega)$ is to be
    continuous.
    In the fit, we will vary $\omega_\mathrm{k}$ in place of $b$ as
    a free parameter.
  \item 
    The sensitivity of the CPLEAR data to the imaginary part of 
    the regeneration amplitude is not sufficient for a parameter fit 
    in the energy range below the first total cross section 
    measurements in the 850 MeV region.
    Following Refs. \cite{Arai} and \cite{Dumbrajs}, we simply 
    assume an essentially linear increase of $\imfm$ from the physical 
    threshold to 830 MeV total laboratory energy, in accordance with the 
    data.
  \item 
    At the physical threshold, we fix $\imfm$ to the value corresponding
    to the measured \PKm\PC\ scattering length.
    The small error on this measurement is negligible in this analysis.
  \item
    Virtually nothing is known about $\imfm(\omega)=\im\;f(\PaKz\PC;\omega,
    \theta=0)$ 
    below the elastic threshold, except that it must vanish at
    $\omega_\PgL$, the beginning of the cut.
    We find that any smooth and small contribution from the integral
    between $\omega_\PgL$ and $m_\PKz$ in (\ref{DR}) is 
    compatible with the experimental information, so that we
    may assume a linear behavior for this energy range as well.
\end{itemize}    
This gives a total of nine parameters to be determined in the fit. 
All of these parameters result from the interpolation of the data 
except for the effective pole, which represents the unphysical region 
below the elastic threshold. 
It turns out, however, that our evaluations are quite insensitive
to the pole position $\omega_\mathrm{p}$. 
   We therefore fix it to the position of the \PgL\ pole, given by
the total energy needed to form a \PCtw$_\PgL$ resonance, 
\begin{equation} \label{pole}
  s = m_{\PC_\PgL}^2 = 
      m_{\PKz}^2 + m_{\PC}^2 + 2\omega_\mathrm{p}m_{\PC}^{} \quad ,
\end{equation}
which gives $\omega_\mathrm{p} =$ 172.3 MeV.

Further parameters are needed for experimental reasons.
As is evident from Fig.~\ref{FigModule}, there are systematic shifts in the 
normalization of the moduli measured by the high energy experiments 
(parameter $\beta$) whereas there is good agreement on the 
slope parameter $\alpha$.
To prevent the systematic normalization uncertainties from affecting
the result on $\alpha$, we allow for three additional correction factors,
$N_\mathrm{Car}$, $N_\mathrm{Alb}$ and $N_\mathrm{Roe}$ for the 
Brookhaven \cite{Carithers77}, Serpukhov \cite{Albrecht, Hladky} and 
Fermilab \cite{Roehrig} experiments, respectively,
to be determined together with the other parameters in the fit.
Since the later Fermilab publication ~\cite{Schwingenheuer} does not contain
individual
data points but only the value of $\alpha$ resulting from their analysis,  
we directly employ this value in our fit.

\section{Fit results and discussion} \label{FitRes}

\begin{figure}[b]
\begin{center}
   \mbox{\epsfig{figure=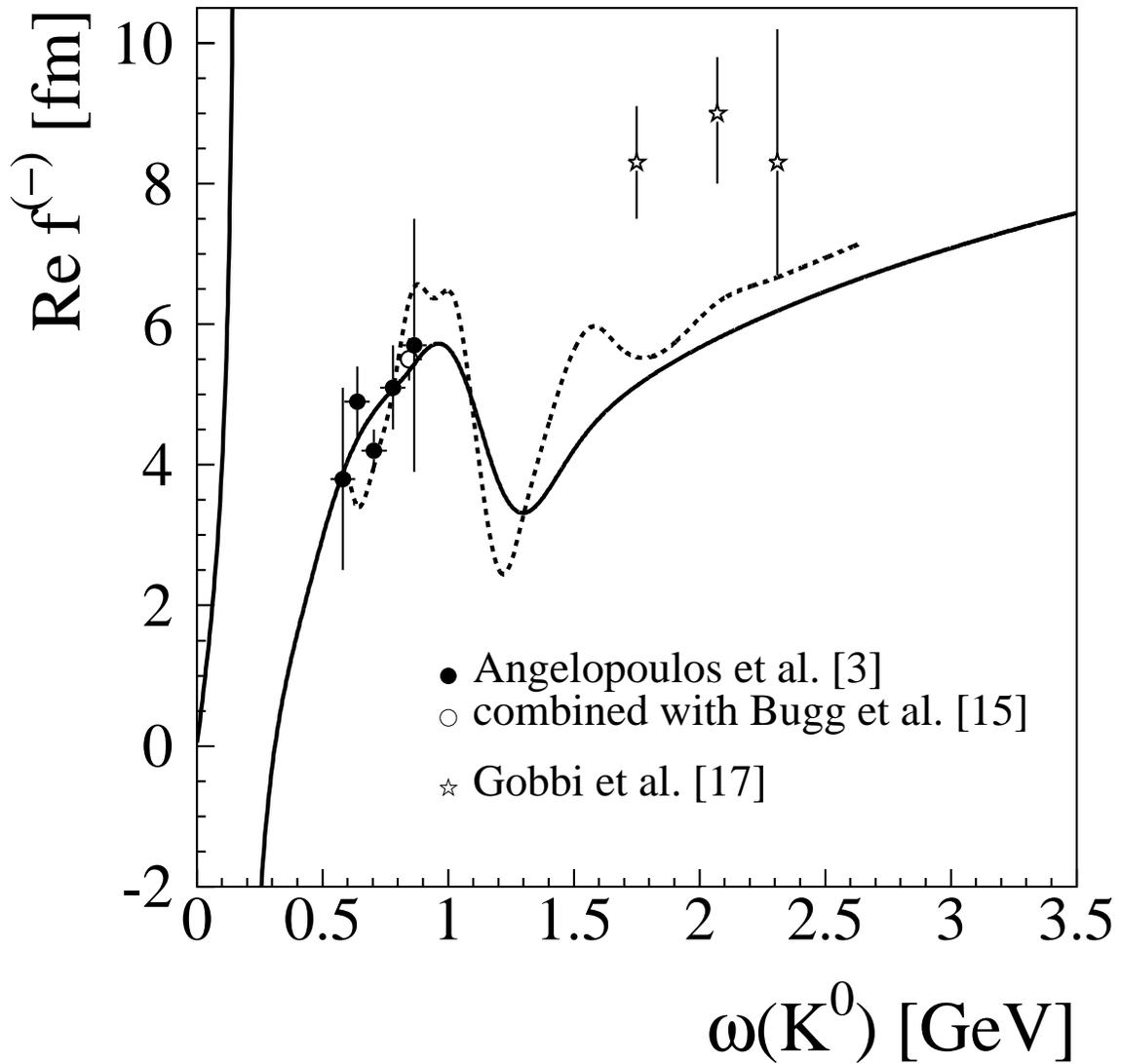,width=16cm}}
\end{center} 
\caption{\label{FigRe} 
             Real part of the regeneration amplitude $\fm$:
             experimental results in comparison with our fit 
             including the pole at $\omega_\mathrm{p} =$ 172.3 MeV 
             (solid line) and the optical model calculation of 
             Ref.~\cite{Eberhard} (dashed line).
             The data of Ref.~\cite{Gobbi} are discarded in the fit.
             Note the linear scale.}
\end{figure}

We now vary the 11 parameters defined in the preceding section and 
compare at each data point the prediction for the real part from (\ref{DR}) 
with the experiment. 
Correlations between different measured quantities are taken into
account where they have been reported, in particular the strong
correlations between our own measurements of imaginary and real parts
\cite{CPLEAR}.
Uncertainties in the kaon beam energy are neglected throughout.
The weighted deviations are then summed up to a $\chi^2$ in the usual way.
Minimization of this $\chi^2$ therefore reflects both the measured imaginary 
and real parts of the amplitude 
and yields the following parameters 
($\chi^2=134.2$ for 127 degrees of freedom):
\begin{eqnarray*}
  r                   &=& -1.57   \pm  0.25                  \\
  C                   &=& (2.55   \pm  0.18)\   \mathrm{fm}  \\
  \omega_\mathrm{res} &=& (1130   \pm  9)\      \mathrm{MeV} \\
  \sigma              &=& (147    \pm  12)\     \mathrm{MeV} \\
  \omega_\mathrm{k}   &=& (4570^{+360}_{-330})\ \mathrm{MeV}  
  \quad\quad
  (\mbox{\rm or}\ b    = (0.10 \pm 0.02)\ \mathrm{fm} )      \\
  a                   &=& 0.61   \pm  0.03                   \\
  \alpha              &=& 0.424  \pm  0.005                  \\
  \beta               &=& (0.35   \pm  0.02)\   \mathrm{fm}  \\
  [0.5\baselineskip]
  N_\mathrm{Car}      &=& 0.96   \pm  0.02                   \\
  N_\mathrm{Alb}      &=& 1.22   \pm  0.07                   \\
  N_\mathrm{Roe}      &=& 1.07   \pm  0.03                   \\
\end{eqnarray*}
The resulting fit functions are shown in Figs.~\ref{FigModule}--\ref{FigRe}, 
together with an optical model calculation \cite{Eberhard}. The results in 
the near-threshold region are only weakly sensitive to the uncertainty in 
the high energy region. In particular, the total contribution of the Regge
asymptotics (\ref{EqRegge}) from $\omega\geq\omega_\mathrm{k}$ to the value
$\refm(m_\PKz)$ is about 0.9 fm, with the estimated error being less than 0.1~fm. 

In order to discuss the physics below the elastic threshold we introduce
the discrepancy function \cite{Ericson}
\begin{equation}   
  \Delta(\omega) = 
  \refm_{\mathrm{exp}}(\omega) - \refm_{\mathrm{phys}}(\omega) 
\label{Deltaomega}
\end{equation}
where $\refm_{\mathrm{phys}}(\omega)$ represents the dispersion integral
(Eq.~\ref{DR}) from the physical region\footnote{%
To avoid a singularity at $\omega=m_\PKz$ in $\fm_{\mathrm{phys}}(\omega)$ 
we use a linear parameterization of $\imfm(\omega)$ for $\omega<m_\PKz$ as 
described in Section~\ref{DRP}. Its contribution to $\Delta(\omega)$ 
is negligible in the energy range of CPLEAR data.
} 
($\omega > m_\PKz$). 
If the unphysical region is well represented in the physical region 
by the effective pole assumed in Eq.~\ref{DR} then the quantity
\begin{equation}  \label{Domega}
   D(\omega) =  \frac{\omega^2-\omega_\mathrm{p}^2}{2\omega}
   \;\Delta(\omega)
\end{equation}
should be independent of the energy and equal to the effective residue\footnote{
More elaborated methods (see \cite{Antolin}) do not help in our case because of 
the scarcity of data.
},
\begin{equation} \label{effRes}
   D(\omega) = r .   
\end{equation}
In Fig.~\ref{FigDelta} we show $D(\omega)$ for 
$\omega_\mathrm{p}$ from Eq.~\ref{pole} corresponding to a 
\PCtw$_\PgL$ system.
The function $D(\omega)$ is not sensitive to the exact pole position and
Fig.~\ref{FigDelta} shows no energy dependence which confirms the
validity of the effective pole ansatz.
The effective residue turns out to be small and negative 
while previous attempts to determine it from FDR led to large and positive 
values:
\begin{equation} 
   r = \left\{ \begin{array}{rcl} 
                 -1.57 \pm 0.25 & \quad &  \mbox{a)\quad this analysis} \\  
                  23.2 \pm 5.8  & \quad &  \mbox{b)\quad Ref. \cite{Arai}} \\
                  12.2 \pm 3.0  & \quad &  \mbox{c)\quad Ref. \cite{Dumbrajs}}
                  \\               
\end{array} \right.   
\label{effResValue} 
\end{equation} 
The main reason for this is that both of these earlier analyses completely 
relied on the data of Ref.~\cite{Gobbi} which we have discarded in view of 
their incompatibility with all other experiments.
Given the large error bars of these data, their inclusion into our fit hardly 
changes the pole residue but increases the $\chi^2$ to 197.6 for now 
133 degrees of freedom.
Without the CPLEAR data for the real part, we would get a large positive 
residue of order of $r=12$, as can be seen from Fig.~\ref{FigDelta}, 
consistent with the earlier analysis \cite{Dumbrajs}. 

Our main result therefore is that the residue is small.  
In our analysis even its negativity is stable against considerable 
variations of a smooth background in the unphysical region and against 
variations of the imaginary part in the physical region below 830~MeV.
However, since the contribution of the effective pole is not large
where the real parts are measured by CPLEAR,
$0.56 \leq \omega \leq 0.9$ GeV, 
the sign is less certain than the magnitude.

\begin{figure}[b]
\begin{center}
    \mbox{\epsfig{figure=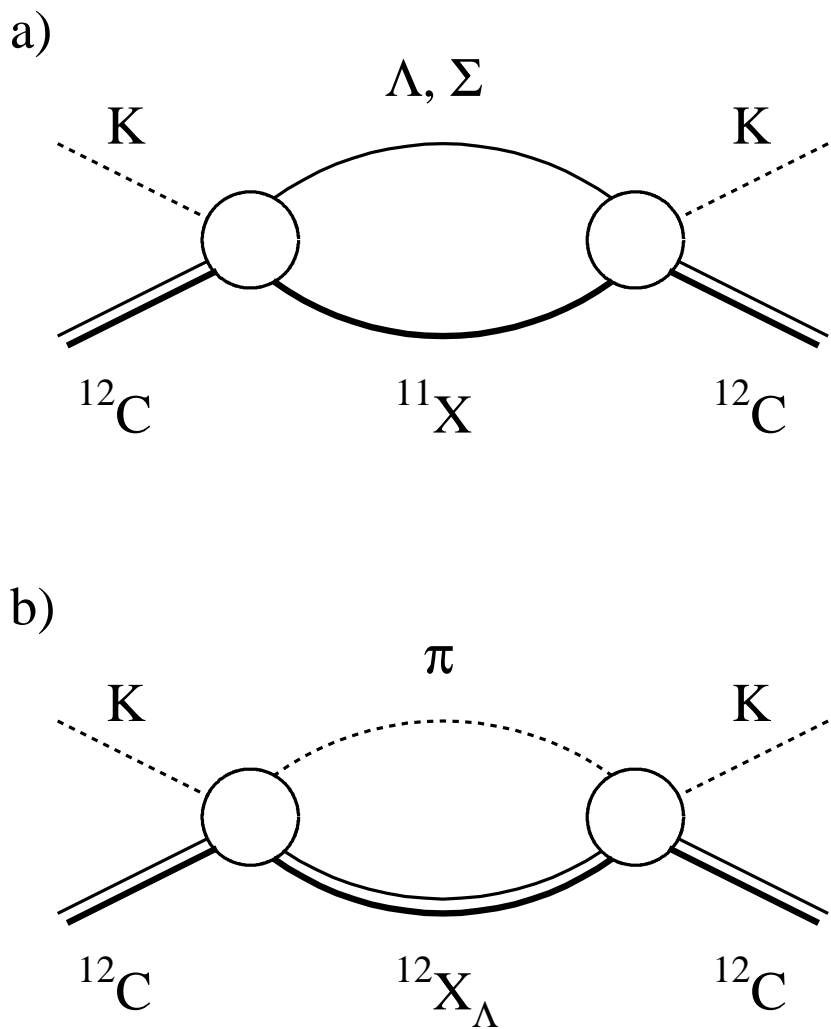,width=16cm}}
\end{center} 
\caption{\label{FigDiag}
         \PK\PCtw\ scattering diagrams for hyperon (a) and 
          pion (b) exchange. The symbol \ensuremath{\mathrm{^{11}X}} 
          denotes a spectator system of 11 nucleons
          (\ensuremath{\mathrm{^{12}X_{\Lambda}}} contains 
          an additional \PgL\ hyperon).}
\end{figure}

\begin{figure}[b]
\begin{center}
   \mbox{\epsfig{figure=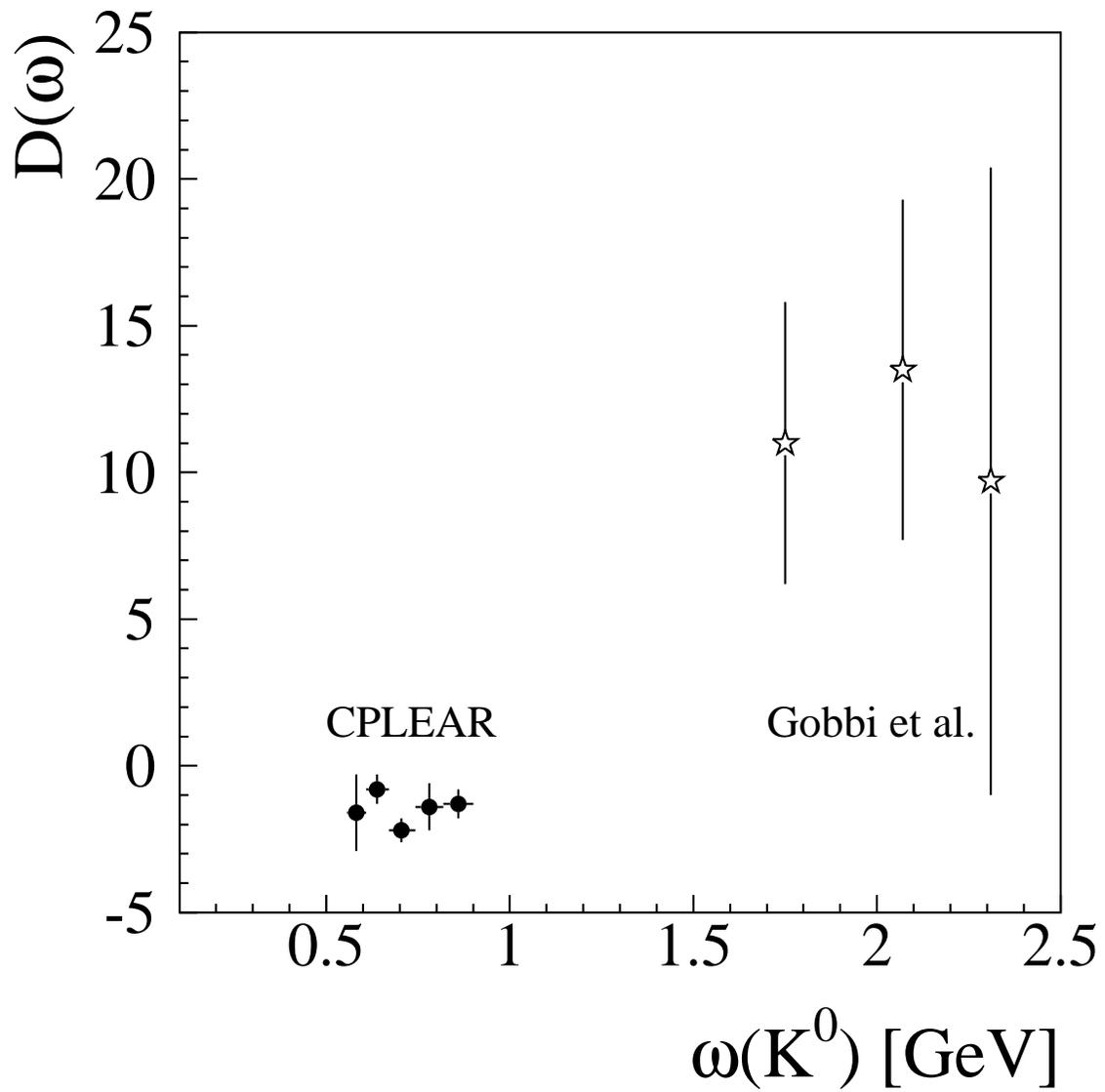,width=16cm}}
\end{center} 
\caption{\label{FigDelta}
             The function $D(\omega)$ as derived from our data
             (full circles) and from the data of Ref.~\cite{Gobbi}
             (see text).}
\end{figure}

From first principles the sign of the residue is directly related to
the parity of the system of the particles exchanged \cite{Ericson}.  
For the elementary \PKpm\Pp\ scattering, hyperon exchange leads to a 
positive residue (similar to \Pgp\PN\ scattering).  
Hence if the unphysical region for \PK\PCtw\ scattering 
were dominated by the superposition of elementary hyperon 
exchange of Fig.~\ref{FigDiag}a, a positive residue would be expected.  
If, however, an additional pion in a relative S-wave is exchanged 
(Fig.~\ref{FigDiag}b) or if the hyperon in Fig.~\ref{FigDiag}a is in 
P-wave relative to the spectator nucleons, 
then the opposite parity results and the residue is negative. 
Our analysis clearly indicates that there are large cancellations between 
these contributions with different parities since the residue is stably 
small. 
   The smallness of the effective residue is actually very gratifying since
the earlier results \cite{Arai,Dumbrajs} claiming a large and positive
residue are very hard to understand. 
   As was already noticed by Dumbrajs \cite{Dumbrajs}, the impulse approximation 
has no chance to work.  He observed that the measured imaginary part (total cross 
section) at low energies are much smaller than the impulse approximation while  
the result for the effective residue 
(\ref{effResValue}c) is about 50\% larger than the sum of the elementary 
residues for $\Lambda$ and $\Sigma$ exchanges. 
The value (\ref{effResValue}b) from \cite{Arai} is even bigger. 
Our result (\ref{effResValue}a) therefore solves a long standing puzzle.

The threshold region deserves some comment.  
In view of the large and repulsive \PKm\PC\ scattering length \cite{Seki}
we have also used the near threshold parameterization of \cite{Arai} 
where this feature is related to a narrow peak in the imaginary part 
below threshold, see Fig.~3. 
This local structure near threshold does not affect our parameters for the
effective pole, as we have checked. 
Our evaluation gives for the real part of the regeneration amplitude at threshold 
$\re\,\fm(m_\PKz)=2.9\;$fm. 
Using the real part of the $\PKm\PC$ scattering length 
$\re\, a_0(\PKm\PC)= 2.9\;$fm from Ref.\cite{Seki} we obtain the real part of 
the $\PKp\PC$ scattering length 
$\re\, a_0(\PKp\PC)= \re\,\fm(m_\PKz) + \re\, a_0(\PKm\PC) = 5.8\;$fm. 
The value of $a_0(\PKp\PC)$ should be considered as an estimate: 
it is sensitive to the parameterization of 
$\imfm(\omega)$ near the elastic threshold because of a steep energy dependence 
of $\refm(\omega)$ (see Fig.\ref{FigRe}). For comparison, using an optical model,  
we obtain $a_0(\PKp\PC)\approx 1.5\;$fm.   
Obviously there is a partial cancellation between the contributions from
\PKm\ and \PKp\ for $\fm$ since both scattering lengths are repulsive.
  While these findings are interesting for the discussion of scattering lengths 
and the detailed behavior of the amplitudes near threshold, these local 
structures are irrelevant for our discussion of the effective pole which 
is dominated by much larger energy scales (the energy range of the 
CPLEAR experiment).

In summary, we conclude that the combined information on the 
$\PK\PCtw$ and $\PaK\PCtw$ amplitudes is very well described by a 
forward dispersion relation containing only a small effective  
pole contribution from the subthreshold region, consistent with 
theoretical expectations. 
   Earlier evaluations claiming a large contribution from the 
unphysical region can be excluded on the basis of the measured regeneration 
amplitude from CPLEAR in combination with the information on this 
reaction from all other consistent experimental data. These results are
of relevance in the analysis of nuclear scattering and of CP-violation
experiments at low energy \cite{dafne}.  

\section*{Acknowledgements}

We thank J.~Ellis and N.E.~Mavromatos for helpful discussions in the
earlier stages of this analysis.
This work was supported by the following institutions: 
the French CNRS/Institut de Physique Nucl\'eaire et de Physique 
  des Particules, 
the French Commissariat \`a l'Energie Atomique, 
the Greek General Secreteriat of Research and Technology, 
the Netherlands Foundation for Fundamental Research on Matter (FOM), 
the Portuguese JNICT,
the Ministry of Science and Technology of the Republic of Slovenia, 
the Swedish Natural Science Research Council, 
the Swiss National Science Foundation, 
the UK Particle Physics and Astronomy Research Council (PPARC), 
and the US National Science Foundation.



\begin{table}[hbt] 
\begin{center}
\caption{Carbon regeneration experiments.}
\vspace{10pt}
\begin{tabular}{lcclll} \hline 
group & Ref.& $p_\PK$ &method& measured          &$\chi^2/\mathrm{dof}$\\
      &     &[GeV/$c$]&      & part of $\fm$& in fit \\ \hline
 &&&&&\\ 
Angelopoulos et al.                  &
\cite{CPLEAR}                        &
0.25--0.75                           &
$\PKz(\PaKz)\rightarrow\Pgpp\Pgpm$   &
Re, Im $^a$                          &
13.1 / 10                            \\

Christenson et al.                   &
\cite{Christenson}                   &
1.1                                  &
$\PKS\rightarrow\Pgpp\Pgpm$          &
mod                                  &
0.1 / 1                              \\

B\"ohm et al.                        &
\cite{Boehm}                         &
2.7                                  &
$\PKS\rightarrow\Pgpp\Pgpm$          &
mod                                  &
1.7 / 1                              \\
 
Bott-Bodenhausen et al.              &
\cite{Bott67}                        &
4.5                                  &
$\PKLS\rightarrow\Pgp\Pgm\Pgn$       &
arg                                  &
1.2 / 1                              \\

                                     &
\cite{Bott69}                        &
                                     &
$\PKLS\rightarrow\Pgpp\Pgpm$         &
mod                                  &
0.0 / 1                              \\

                                     &
                                     &
                                     &
                                     &
arg                                  &
0.8 / 1                              \\

Carithers et al.                     &
\cite{Carithers75a}                  &
4--10                                &
$\PKLS\rightarrow\Pgp\ell\Pgn$       &
arg                                  &
13.0 / 13                            \\

                                     &
\cite{Carithers75b}                  &
                                     &
$\PKLS\rightarrow\Pgpp\Pgpm$         &
arg $^b$                             &
7.9 / 6                              \\

                                     &
\cite{Carithers77}                   &
                                     &
                                     &
mod                                  &
6.4 / 6                              \\

Hladky et al.                        &
\cite{Hladky}                        &
10--30                               &
$\PKLS\rightarrow\Pgpp\Pgpm$         &
mod                                  &
1.2 / 5                              \\

                                     &
                                     &
                                     &
                                     &
arg                                  &
0.0 / 1                              \\

Albrecht et al.                      &
\cite{Albrecht}                      &
16--40                               &
$\PKLS\rightarrow\Pgpp\Pgpm$         &
mod                                  &
1.3 / 6                              \\

                                     &
                                     &
                                     &
                                     &
arg                                  &
0.4 / 3                              \\

Roehrig et al.                       &
\cite{Roehrig}                       &
30--130                              &
$\PKLS\rightarrow\Pgpp\Pgpm$         &
mod                                  &
21.6 / 10                            \\

                                     &
                                     &
                                     &
                                     &
arg                                  &
9.0 / 10                             \\

Schwingenheuer et al.                &
\cite{Schwingenheuer}                &
20--160                              &
$\PKLS\rightarrow\Pgpp\Pgpm$         &
mod, arg $^c$                        &
0.3 / 1                              \\

\hline
\end{tabular} 
\end{center}
$^a$ {\footnotesize
  Likelihood plots in the complex plane are given.}\\
$^b$ {\footnotesize Published data are analyzed using the value of
  $\phi_{+-}$ resulting from the compilation of Ref. \cite{PDG}.}\\
$^c$ {\footnotesize
  Only a fitted value for the power law exponent $\alpha$ is published.}
\\

\end{table}

\begin{table}[hbt] 
\begin{center}
\caption{Carbon \PKpm\ experiments.}
\vspace{10pt}
\begin{tabular}{lcclll} \hline 
group & Ref.& $p_\PK$ & measured   & derived    &$\chi^2/\mathrm{dof}$\\
      &     &[GeV/$c$]& quantities & part of $\fm$&in fit     \\ \hline
 &&&&&\\ 

Bugg et al.                          &
\cite{Bugg}                          &
0.6--2. 6                            &
$\sigma_\mathrm{T}(\PKpm)$           & 
Im                                   &
17.5 / 9                             \\

Abrams et al.                        &
\cite{Abrams}                        &
1.0--3.3                             &
$\sigma_\mathrm{a}(\PKpm)$           & 
Im                                   &
37.8 / 41                            \\

Gobbi et al.                         &
\cite{Gobbi}                         &
1.68--2.26                           &
$\sigma_\mathrm{T}(\PKpm)$           &
Im                                   &
(discarded)                          \\

                                     &
                                     &
                                     &
$\d\sigma/\d q^2(\PKpm)$             &
Re                                   &
(discarded)                          \\

Afonasyev et al.                     &
\cite{Afonasyev}                     &
1.8                                  &
$\sigma_\mathrm{T}(\PKpm)$           & 
Im                                   &
0.8 / 1                              \\

Backenstoss et al.~/                 &
\cite{Backenstoss}                   &
0                                    &
$a_0(\PKm)$                          & 
Im                                   &
(fixed)                              \\

\quad Seki                           &
\cite{Seki}                          &
                                     &
                                     &
                                     &
                                     \\ 
\hline
\end{tabular} 
\end{center}
\end{table}


\end{document}